\documentclass[epj,twocolumn]{webofc}
\usepackage[varg]{txfonts}   
\woctitle{ISVHECRI 2016}
\newcommand{\rr}{\mathbf{r}}
\newcommand{\intgr}{\int\limits_0^{\infty}}

\begin{document}
\title{Average value of the cosmic ray injection exponent at  Galactic sources}

\author{Anatoly Lagutin\inst{1}\fnsep\thanks{\email{lagutin@theory.asu.ru}} \and
        Nikolay Volkov\inst{1}\fnsep\thanks{\email{volkov@theory.asu.ru}}
}

\institute{Altai State University, Radiophysics and Theoretical Physics Department, 656049, 61 Lenin ave, Barnaul, Russia 
          }

\abstract{%

The energy spectrum of high energy cosmic rays emitted by sources change during propagation in the Galaxy. Using the spectrum observed at the Earth and two different transport models, based on anomalous and normal diffusion equations, we have retrieved an average value of the cosmic ray injection exponent at  Galactic sources taking into account the inhomogeneity of the interstellar medium.  We have shown that the average value of the injection index $p$, obtained in the framework of both transport models, equals to $p\sim (2.8-3.0)$.
}

\maketitle

\section{Introduction}\label{intro}

The spectrum of cosmic rays (CRs)  measured at the Earth is different from their source spectrum. A key to understanding this difference is the determination of how cosmic ray (CR)  particles propagate through the turbulent interstellar
medium (ISM). 

Assuming that the medium is quasi-homogeneous, the propagation process is described by  Ginzburg-Syrovatsky's normal diffusion equation~\cite{Ginzburg:1964,Berezinskii:1990}

\begin{equation}~\label{GSeq}
\frac{\partial N(\rr,t,E)}{\partial t}= D(E)\Delta N(\rr,t,E)+ S(\rr,t,E),
\end{equation}
where $N(\rr,t,E)$ is the density of particles with energy $E$ at the location
$\rr$ and time $t$, $D (E)=D_0E^{\delta}$ is the diffusion coefficient assumed to be spatially constant and $S(\rr,t,E)$ is the source term.

The steady state solution of  ~\eqref{GSeq}, frequently used for interpretation of the CR phenomena, has the form 
\begin{equation}~\label{GSstea}
N(\rr,E) \sim E^{-p -\delta},
\end{equation}
where $p$ is the injection spectrum exponent at the Galactic sources. 
The same result may be obtained using the entirely uniform Leaky-Box model.

The steady state energy spectrum  ~\eqref{GSstea} leads to a very simple  rule to retrieve the exponent $p$ from the  spectrum $ N(E) \propto E^{-\eta}$ observed at the Earth: $p = \eta - \delta$. Since $\eta \approx 2.7$, for $\delta \sim 0.3$ we find that the slope of the CR injection spectrum at the source is $\sim 2.4$. 

However, recent results show that homogeneous diffusion models, based on~\eqref{GSeq}, failed to reproduce observations (see, for example,~\cite{Johannesson:2016}). Multiscale  structures in the Galaxy, found during the last few decades, may be taken as a support to this conclusion. 

Theory and observations show that the ISM is inhomogeneous (fractal-like) on the hundreds of parsecs
scales~\cite{Elmegreen:2004,Bergin:2007,Sanchez:2008}. Stars formation regions also demonstrate fractal features with  spatial scales up to about a kpc~\mbox{\cite{Efremov:2003,Fuente:2009,Sanchez:2010}}.

The particles emitted by Galactic sources
en route to the Solar system pass through  regions of the Galaxy that have different properties.
In such a inhomogeneous ISM, the normal diffusion model is certainly  not kept valid. 

The main goal of this paper is to retrieve the cosmic ray injection
exponent at the Galactic sources from the spectrum observed at Earth taking into account the inhomogeneity of the turbulent interstellar medium. Two different models, based on anomalous and normal  diffusion equations, are used to relate the spectrum injected by the source and that measured at Earth.

In the following section we give an overview of the anomalous diffusion model, discuss our injection exponent retrieval technique and the results obtained in this model. In section 3 we present a new approach to describe     
cosmic ray propagation in the non-homogeneous ISM, based on the normal  diffusion equation~\eqref{GSeq}. Our findings and conclusions are given in section 4. 

\section{Retrieval of the injection exponent  using the anomalous diffusion model}

The non-homogeneous character of matter distribution and associated magnetic field leads to the need to incorporate these ISM features into the cosmic ray diffusion model. A possible way to generalize the  normal diffusion model is to replace the assumption about statistical homogeneity of inhomogeneities distribution by their fractal distribution. An important consequence of this generalization is the power-law distribution of free paths, $r$, in such a medium $p(\rr,E) \propto A(E,\alpha)r^{-\alpha - 1}, r \rightarrow \infty,  0 < \alpha < 2$ --- so-called L\'{e}vy flights.
Since an intermittent magnetic field of the fractal-like ISM leads to nonzero probability of a long stay of particles in inhomogeneities, the presence of the L\'{e}vy trap can not be excluded. In the general case, the probability density function $q(t,E)$ of time, $t$, during which a particle is trapped in the inhomogeneity (L\'{e}vy trap), also has a power-law behavior: $q(t,E) \propto B(E,\beta)t^{-\beta - 1},   t \rightarrow \infty,  \beta < 1$. 
 
A generalization of the homogeneous normal diffusion model to the case of inhomogeneous (fractal-like) ISM, has been made for the first time in our papers~\cite{Lagutin:2000,Lagutin:2001}. Later, it was shown~\cite{Lagutin:2003,Lagutin:2003ew,LagutinTymentcev:2004,Lagutin:2005ijmp,Lagutin:2009,Volkov:2015}   that an anomalous cosmic ray diffusion model, developed by the authors, allows to describe the main features of  nuclei, electron and positron spectra observed in the Solar system. Particularly, in the anomalous diffusion model the key feature of the all particle energy spectrum --- the knee at $3\cdot 10^{15}$~eV ---  appears naturally without additional assumptions.

The equation for the density of particles with energy, $E$, at the location, $\rr$, and time, $t$, generated in a fractal-like medium by Galactic sources with a distribution density $S(\rr,t,E)$ can be written as~\cite{Lagutin:2003,LagutinTymentcev:2004}
\begin{multline}~\label{SuperEq}
\frac{\partial N(\rr,t,E)}{\partial t}=-D(E,\alpha,\beta)\mathrm{D}_{0+}^{1-\beta}(-\Delta)^{\alpha/2} N(\rr,t,E)+\\ + S(\rr,t,E).
\end{multline}
Here $\mathrm{D}_{0+}^{\mu}$ denotes the Riemann-Liouville fractional derivative~\cite{Samko:1993} and $(-\Delta)^{\alpha/2}$ is the fractional Laplacian (`Riesz operator')~\cite{Samko:1993}. The anomalous diffusion coefficient $D(E,\alpha,\beta) \sim A(E,\alpha)/B(E,\beta) = D_0(\alpha,\beta)E^{\delta}$.

\begin{figure}[t!]
\centering

\sidecaption

\includegraphics[width=.35\textwidth,angle=-90]{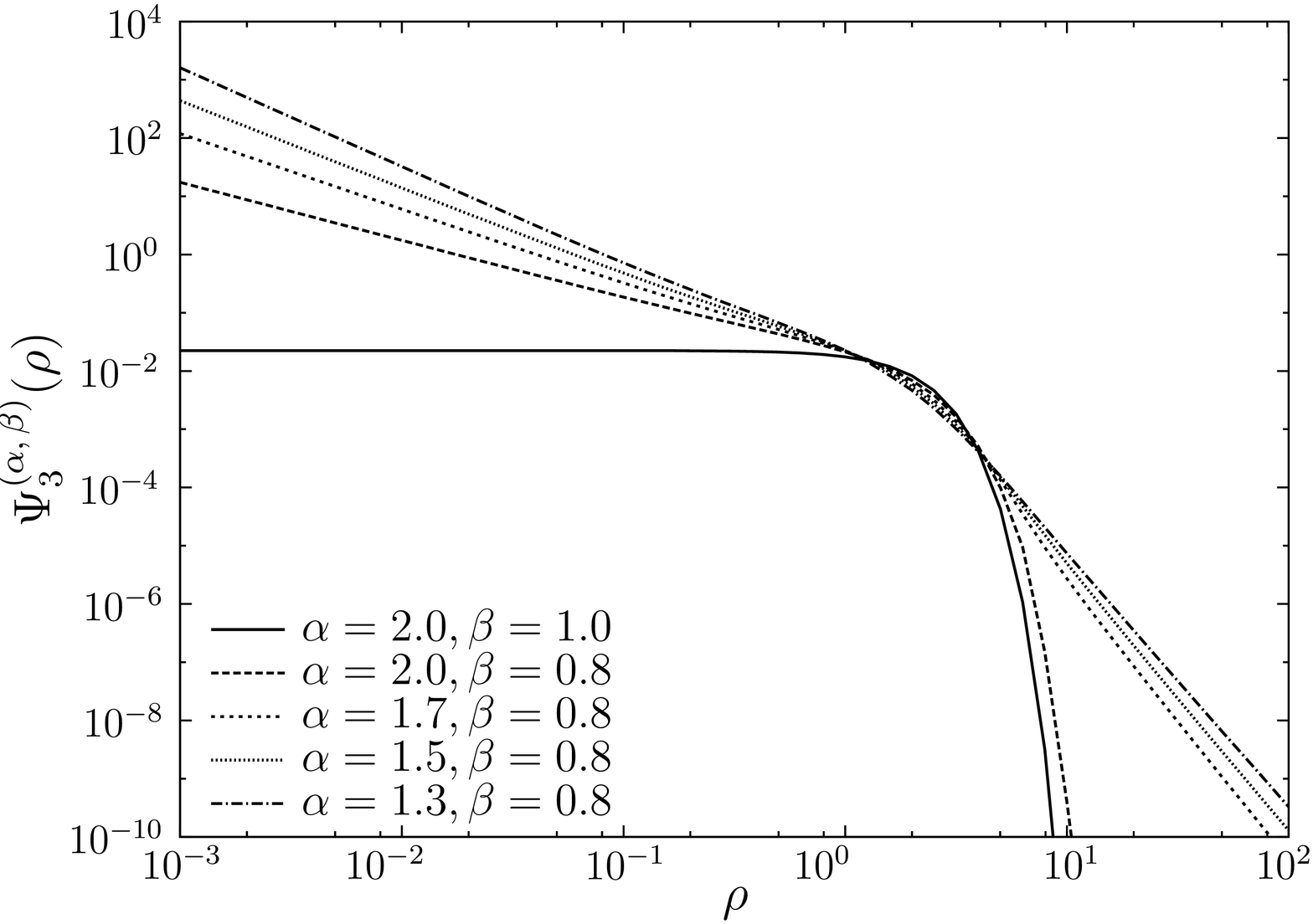}

\caption{Probability density of fractional stable distribution $\Psi_3^{(\alpha,\,\beta)}(\rho)$ for different parameters $\alpha$ and $\beta$}

\label{fig:psi}

\end{figure}

The solution of Eq.~\eqref{SuperEq} for a  point instantaneous source with a 
power-law injection spectrum $S(\rr,t,E)=S_{0} E^{-p}\delta(\rr) \delta(t)$ has the form~\cite{Lagutin:2003,LagutinTymentcev:2004}
\begin{multline*}\label{eq:solanomdifeq}
N(\rr,t,E) = S_{0} E^{-p}(D(E,\alpha,\beta)t^{\,\beta})^{-3/\alpha}\times\\
\times\Psi_3^{(\alpha,\,\beta)}(|\rr|(D(E,\alpha,\beta)t^{\,\beta})^{-1/\alpha}),
\end{multline*}
where $\Psi_3^{(\alpha,\,\beta)}(\rho)$ is the density of the fractional stable distribution~\cite{Uchaikin:1999a,Zolotarev:1999}
\begin{equation*}
  \Psi_3^{(\alpha,\,\beta)}(\rho)=\int\limits_0^\infty{g_3^{(\alpha)}({r\tau^\beta})q_1^{(\beta,1)}(\tau)\tau^{3\beta/\alpha}d\tau}.
\vspace*{-3mm}
\end{equation*}
Fig.~\ref{fig:psi} demonstrates the behaviour of the function $\Psi_3^{(\alpha,\,\beta)}(\rho)$ for different parameters $\alpha$ and $\beta$. 

\begin{figure}[!t]
\centering

\includegraphics[width=.35\textwidth,angle=-90]{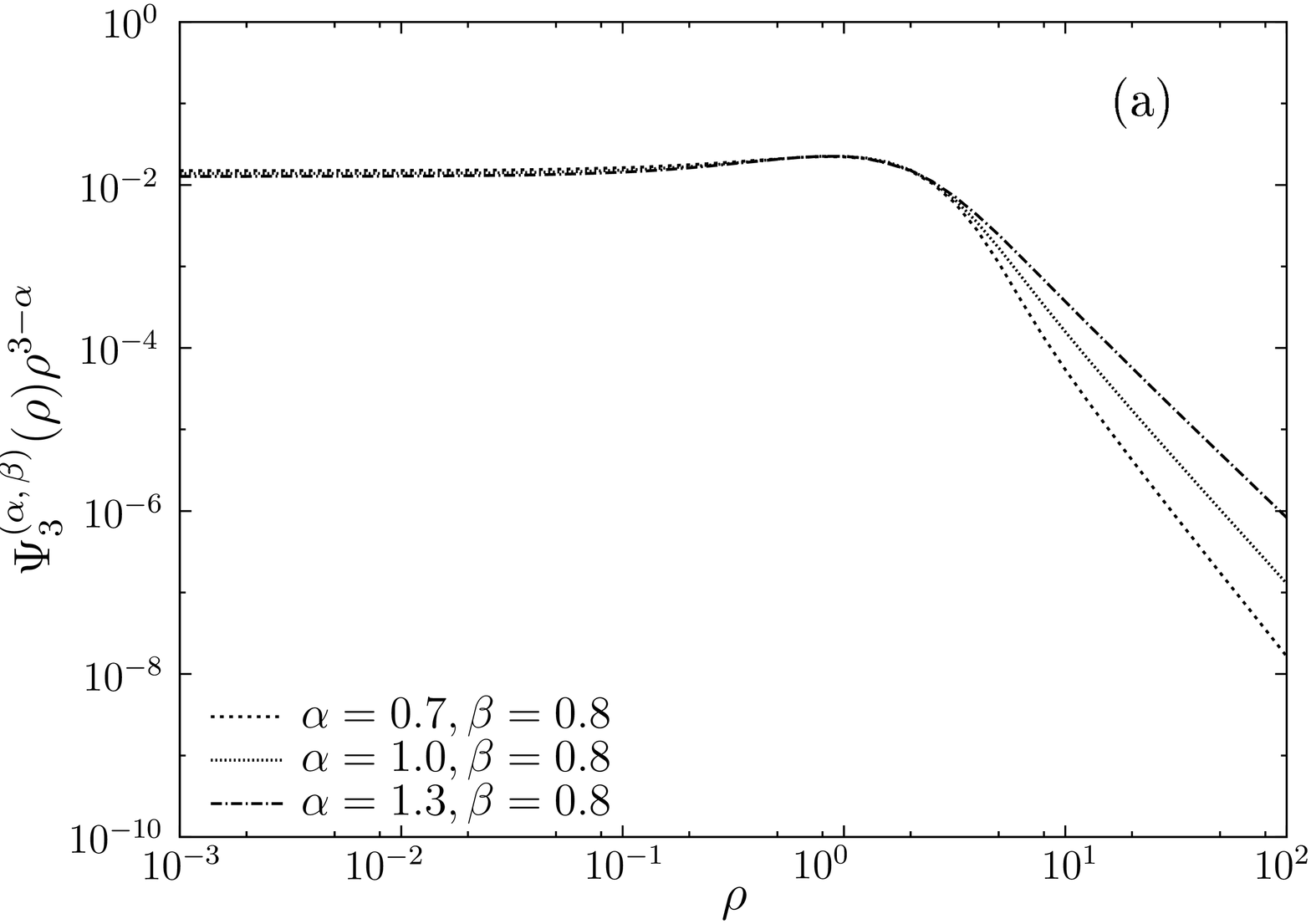}

\sidecaption

\includegraphics[width=.35\textwidth,angle=-90]{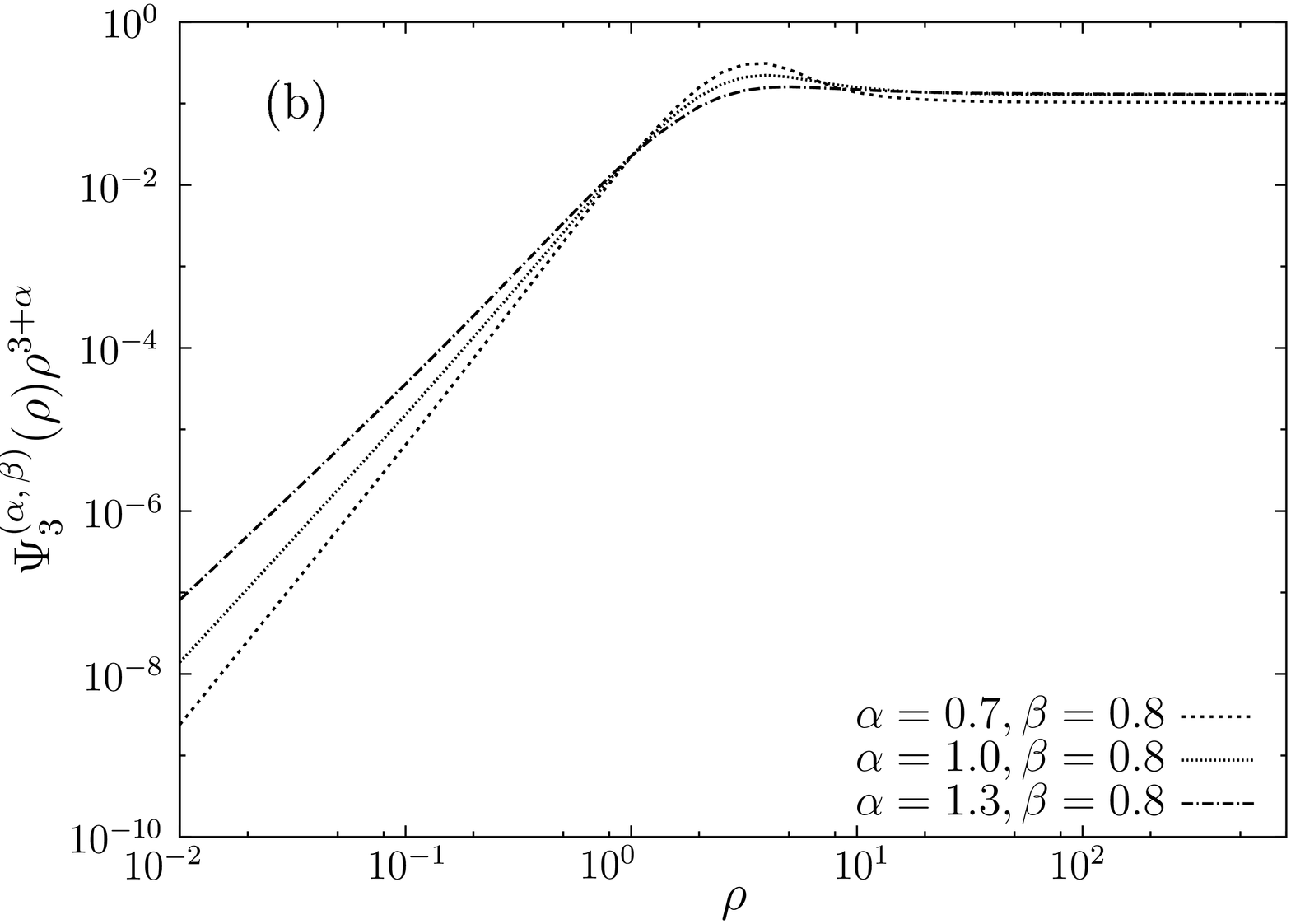}

\caption{Asymptotic behavior of the function $\Psi_3^{(\alpha,\,\beta)}(\rho)$ for \mbox{$\rho\rightarrow 0$} (a) and for $\rho\rightarrow\infty$ (b)}\label{fig:psiass}
\end{figure}

The asymptotic behaviour of the scaling function $\Psi_3^{(\alpha,\,\beta)}(\rho)$ for $\alpha < 2, \beta < 1$,
$$\Psi_3^{(\alpha,\,\beta)}(\rho) \propto \rho^{-(3-\alpha)},\quad \rho\rightarrow 0,$$
$$\Psi_3^{(\alpha,\,\beta)}(\rho) \propto \rho^{-3-\alpha},\quad \rho\rightarrow\infty,$$
is shown on Fig.~\ref{fig:psiass}. 

Taking into account that  $\rho \equiv |\rr|(D(E,\alpha,\beta)t^{\,\beta})^{-1/\alpha}$, we find
\begin{equation*}
N \sim E^{-\eta} = E^{-p + \delta}, \quad E \ll E_k 
\end{equation*}
and
\begin{equation*}
N \sim E^{-\eta} = E^{-p - \delta}, \quad E \gg E_k, 
\end{equation*}
where $E_k$ is the knee energy. 
Since $\Delta\eta=\eta|_{>E_k} - \eta|_{<E_k}$ is known from experimental data to  equal $\Delta\eta \sim0.6$, the last equations permit to,  self-consistently, retrieve  both spectral exponents $p$ and $\delta$:
$$\delta= \Delta\eta/2 \sim 0.3,$$
$$p = \eta|_{<E_k} + \delta = \eta|_{>E_k} - \delta  \approx 2.8\div2.9.$$   
At $E = E_k$ the spectral index $\eta(E_k)$  is equal to the injection exponent $p$ at  the Galactic source.

Note that in the energy range below the knee the cosmic ray density 
is 
\begin{equation}\label{eq:anomdiffresult}
N(\rr,t,E) \sim \dfrac{E^{-p+\delta}}{r^{3 +\alpha}}. 
\end{equation}

\section{Normal diffusion of cosmic rays in a non-homogeneous ISM}\label{sec:newapproach}

The medium properties of an inhomogeneous ISM vary in space and time. The particles emitted by Galactic sources en route to the Solar system pass through  regions of the Galaxy that have  different diffusion coefficients. 
 
Here we suppose that the particles with probability density  $\varphi(D_0)$ propagate in the ISM with a diffusion coefficient $D_0$. Since for a homogeneous ISM, with the constant diffusion coefficient $D_0$, the solution of a normal diffusion equation~\eqref{GSeq} for a  point  instantaneous source with a power-law injection spectrum has the  form
\begin{equation}\label{eq:gssol}
N(\rr,t,E) = \dfrac{S_0 E^{-p}}{(4\pi D_0E^{\delta} t)^{3/2}}\exp\left(-\dfrac{r^2}{4D_0E^{\delta}t}\right),
\end{equation}
we can find the density of particles visiting this part of the Galaxy,
\begin{equation*}
dN_D(\rr,t,E) = \dfrac{S_0E^{-p}\varphi(D_0)dD_0}{(4\pi D_0E^{\delta} t)^{3/2}}\exp\left(-\dfrac{r^2}{4D_0E^{\delta}t}\right),
\end{equation*}
and then the observed density $N(\rr,t,E)$
\begin{multline}\label{eq:nnener}
N(\rr,t,E) = \int dN_D(\rr,t,E)  = \dfrac{S_0E^{-p-3\delta/2}}{(4\pi t)^{3/2}}\times\\[5mm]
\times\intgr\dfrac{1}{D_0^{3/2}}\exp\left(-\dfrac{r^2}{4D_0E^{\delta}t}\right)\varphi(D_0)dD_0.
\end{multline}

In order to have insight into possible types of diffusivity distribution
we follow the approach presented in~\cite{Petrovskii:2009}. If $r$ is the random jump length of a particle and $\tau$ is the time required to make this jump, the diffusion coefficient is given by equation
$$D_0=\frac{r^2}{2\tau}.$$ 
Since $r/\tau$ is the speed $v$, we have   
$$D_0=\frac{v^2\tau}{2}.$$
Assuming that the speed of particle motion $v$ along its trajectory is the same for all particles, for $\varphi(D_0)$ we find
\begin{equation*}  
\varphi(D_0) = \psi(\tau)\left|\dfrac{d\tau(D_0)}{dD_0}\right|= \frac{2}{v^2}\psi(\tau).
\end{equation*}
It is seen, for example, that the assumption that the time, $\tau$, of a single step is distributed normally leads to a normal distribution for diffusivity. In the case of $\psi(\tau) \propto \tau^{-\beta - 1}$, \mbox{$\beta < 1$} the power-law distribution for $\varphi(D)$ may be obtaned. 

Similar distributions may be obtained if we assume that all particles during the random walk have the same value of $r$. In this case $\tau(D_0) =r^2/(2D_0)$ and
\begin{equation}\label{eq:gs2}  
\varphi(D_0) = \psi(\tau)\left|\dfrac{d\tau(D_0)}{dD_0}\right|= \frac{r^2}{2{D_0}^2}\psi(\tau).
\end{equation}
Using, for example, a normal distribution for $\tau$
\begin{equation*}
\psi(\tau) = \dfrac{1}{\sqrt{\pi}\Delta\tau}\exp\left[ -\left(\dfrac{(\tau - \tau_0)}{\Delta\tau}\right)^2\right],
\end{equation*} 
where $\tau_0$ and $\Delta\tau$ are parameters ($\Delta\tau \ll \tau_0$), for large $D_0$ from~\eqref{eq:gs2} we find 
$$ \varphi(D_0) \sim {D_0}^{-2}. $$
The probability distribution function $\varphi(D_0)$ has a power-law form.

Here we consider the power-law distribution $\varphi(D_0)\sim D_0^{-\gamma}$ for $D_0$ $\rightarrow\infty$, where $\gamma$ is a positive parameter. Since $\varphi(D_0)$ has the meaning of a
probability density, that is  $\int\limits_0^\infty\varphi(D_0)dD_0 = 1$, we must assume that $\gamma>1$.

In order to calculate the integral in equation~\eqref{eq:nnener}, we need to make some additional assumptions about the properties of $\varphi(D_0)$  at small and also intermediate values 
of $D_0$. We adopt the following functional form
\begin{equation}\label{eq:fi}
\varphi(D_0) = A D_0^{-\gamma} \exp\left(-\dfrac{\mathfrak{D}}{D_0}\right),
\end{equation}
where $A=\dfrac{\mathfrak{D}^{\gamma-1}}{\Gamma(\gamma-1)}$ is the scaling factor, $\mathfrak{D}$ is the characteristic diffusivity for the Galactic medium, $\gamma >1$ is an auxiliary parameter, $D_0>0$. We additionally define $\varphi(0)=0$.

Having substituted equation~\eqref{eq:fi} into equation~\eqref{eq:nnener} we find
\vspace{-1mm}
\begin{multline}\label{eq:nnnener}
N(\rr,t,E) = \dfrac{S_0 A E^{-p-3\delta/2}}{(4\pi t)^{3/2}}\times\\
\times\intgr\dfrac{dD_0}{D_0^{\gamma+3/2}}\exp\left[-\left(\mathfrak{D}+\dfrac{r^2}{4E^{\delta}t}\right)\dfrac{1}{D_0}\right].
\vspace{3mm}
\end{multline}
Using new variables $p=\mathfrak{D}+\dfrac{r^2}{4E^{\delta}t}$ and
$D_0=\dfrac{p}{z}$, the integral in~\eqref{eq:nnnener} is transformed as follows:
\begin{multline*}
\intgr D_0^{-\gamma-3/2}\exp\left(-\dfrac{p}{D_0}\right)dD_0 = \\ = -p\int\limits_{\infty}^0 \dfrac{p^{-\gamma-3/2}}{z^{-\gamma-3/2}}\exp(-z)\dfrac{dz}{z^2} = \\
= p^{-\gamma - 1/2} \intgr z^{\gamma - 1/2} \text{e}^{-z} dz = p^{-\gamma - 1/2}\Gamma\left(\gamma+\dfrac{1}{2}\right).
\end{multline*}
As a result, we obtain

\begin{equation}\label{eq:nitener}
N(\rr,t,E) = \dfrac{S_0 AE^{-p-3\delta/2}}{(4\pi t)^{3/2}}\left(\mathfrak{D}+\dfrac{r^2}{4E^{\delta}t}\right)^{-\gamma-1/2}\Gamma\left(\gamma+\dfrac{1}{2}\right).
\vspace{2mm}
\end{equation}

At any given moment of time, $t$, and for sufficiently large $r$, so that $r^2 \gg 4 \mathfrak{D} E^\delta t$, the density of particles, $N(\rr,t,E)$,  has the form
\begin{equation*}
N(\rr,t,E) = \dfrac{S_0\mathfrak{D}^{\gamma-1}}{\pi^{3/2}(4t)^{1-\gamma}}\dfrac{\Gamma\left(\gamma+\frac{1}{2}\right)}{\Gamma\left(\gamma-1\right)}\dfrac{E^{-p-\delta+\delta\gamma}}{r^{2\gamma+1}}
\end{equation*}
or
\begin{equation}\label{eq:asimpt}
N(\rr,t,E) \sim \dfrac{E^{-p-\delta+\delta\gamma}}{r^{2\gamma+1}}.
\end{equation}

Thus, it can be seen, that the function $\varphi(D_0)$ given by equation~\eqref{eq:fi} leads to an anomalous spatial distribution of the particle density. Instead of the Gaussian distribution of particle density~\eqref{eq:gssol}, which is predicted by the homogeneous normal diffusion models, the large-distance asymptotical behavior, we have obtained, is described by a power law $\sim {r^{2\gamma+1}}$. 

In the discussed model~\eqref{eq:nitener} the energy spectrum 
observed at Earth is $ N(E) \propto E^{-\eta} = E^{-p-\delta+\delta\gamma}$. From the slope   $\eta = p + \delta(1 - \gamma$) we can find the injection spectrum exponent, $p$, at the Galactic source of the cosmic rays: $p =\eta -\delta(1 - \gamma$). It differs from the steady state energy spectrum exponent $p = \eta - \delta$. Since  $\gamma >1$, we have $p > \eta$.

It should be noted that a similar anomalous spatial distribution of the particle density may be found in the case of $\varphi(D_0) \sim D_0^{\theta} \exp \left(-\dfrac{D_0}{\mathfrak{D}}\right)$, where $\mathfrak{D}$ is the characteristic diffusivity for the Galactic medium and $0<\theta<1$.  This distribution describes a model with  exponential decrease at $D_0 \rightarrow \infty$.  
For this scenario we have found
\begin{equation*}
N(\rr,t,E) \sim r^{\theta -1}\exp\left( -\dfrac{r}{\sqrt{\mathfrak{D}E^{\delta}t}}\right) E^{-p-\delta(\theta +1)/2}.
\end{equation*}

\section{Conclusions}

We have performed the retrieval of the cosmic ray injection
exponent at the Galactic sources from the spectrum observed at Earth taking into account the  inhomogeneity of the turbulent interstellar medium. Two different transport models, based on anomalous and normal  diffusion equations, have been used to relate the spectrum injected by the source with that measured at Earth. 

Using the normal diffusion model, we have shown that the variation of the ISM properties leads to an anomalous spatial distribution of the particle density. Its asymptotic behavior~\eqref{eq:asimpt}  for  $1.5 < \gamma < 2$ practically coincides with our result~\eqref{eq:anomdiffresult} obtained in the framework of the anomalous diffusion model for $1<\alpha<2$. In other words, the cosmic rays observed at Earth collectively appear to display non-diffusive (superdiffusive) characteristics although an individual particle has moved in a diffusive manner.

We have shown that the average value of the injection index $p$, obtained in the framework of both transport models, equals  $p\sim (2.8\div 3.0)$.

We note that a similar steep spectrum of accelerated particles has been observed from supernova remnants 
W44~\cite{Abdo:2010S} and IC 443~\cite{Abdo:2010A} with the Fermi Large Area Telescope. The value $p=3$ has been found in~\cite{Tanaka:2008}  for RX J1713.7-3946.

\section*{Acknowledgements}
The authors acknowledge support from The Russian Foundation for Basic Research
grant No. 16-02-01103.

\end{document}